\documentclass[aps,12pt, a4paper,nofootinbib]{revtex4}

\usepackage[brazil, english]{babel}
\usepackage[utf8]{inputenc}
\usepackage[T1]{fontenc}
\usepackage{amsmath}
\usepackage{amsfonts}
\usepackage{amssymb}
\usepackage{graphics,graphicx}
\usepackage{ulem}
\usepackage{graphicx,color}
\usepackage{subfigure} %colocar figura lado a lado, figura a) e b).
\usepackage{wrapfig} % pacote reponsavel para colocar figura ao lado do texto
\usepackage{epstopdf}
\graphicspath{{figuras/}}

\usepackage{setspace}
\usepackage[unicode=true,bookmarks=false,breaklinks=false,pdfborder={0 0 1},colorlinks=true]
 {hyperref}
\hypersetup{
 citecolor=blue,linkcolor=blue,urlcolor=blue}
\begin{document}

\title{Entanglement Entropy in Horndeski Gravity}
%\end{center} 
\author{F. F. Santos$^{1}$}
\email{fabiano.ffs23@gmail.com}  
\address{$^{1}$Instituto de F\'isica, Universidade Federal do Rio de Janeiro, 21941-972 Rio de Janeiro, Brazil.}
\date{\today}

\begin{abstract}
In this work, we explore the holographic entanglement entropy with  an infinite strip region of the boundary in Horndeski gravity. In our prescription we consider the spherically and planar topologies black holes in the AdS$_{4}$/CFT$_{3}$ scenario. In such framework, we show the behavior of the entanglement entropy in function of the Horndeski parameters. Such parameters modify the information store of subsystem A, especially when the parameter $\gamma$ increases the information about the subsystem will also increase or decrease when it decreases. Thus, with this scheme we compute the “first law of entanglement thermodynamics” in Horndeski gravity and we show that a very small subsystem obeys the analogous property of the first law of thermodynamics if we excite the system.
\end{abstract}

\maketitle
\newpage

%\tableofcontents
\newpage
\section{Introduction}

\begin{figure}[!ht]
 \centerline{\includegraphics[scale=0.9]{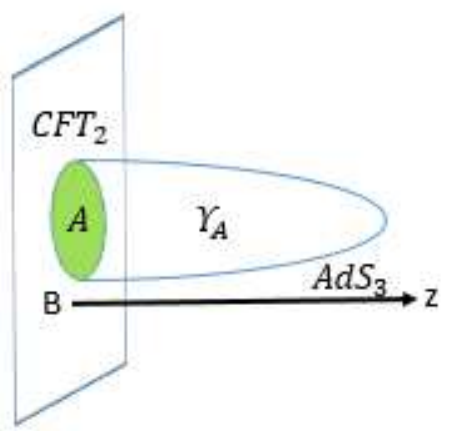}}
\caption{The calculation of holographic entanglement entropy.}\label{3.0}
\label{planohwkhz}
\end{figure}

In recent years, the scenario of entanglement entropy have call attention in special the works of \cite{Takayanagi:2012kg,Bhattacharya:2019zkb,Bhattacharya:2017gzt,Bhattacharya:2012mi,Calabrese:2004eu,Holzhey:1994we,Ryu:2006ef,Ryu:2006bv,Susskind:1994sm,Chaturvedi:2016kbk,Tonni:2010pv,Mansoori:2015sit,Caputa:2013lfa,Blanco:2013joa,Susskind:2021esx,Kay:2016kbi,Park:2015afa,He:2014lfa} where the $S_{A}$ as the entropy for an observer accessible only to the $A$ subsystem and it cannot receive any $B$ signal. Such subsystem $B$ is analogous to the interior of the black hole horizon. For an observer who is sitting on $A$, that is, outside the horizon. On the other hand, because this analogy is not correct, the quantum correction involving a loop for the entropy BH in the presence of fields of the matter is known to be equal to the entanglement entropy \cite{Susskind:1994sm}. Thus, we have that this interesting relation provides us with an important tip to find the holographic dual of the entanglement entropy. In the context of the AdS$_{3}$/CFT$_{2}$ correspondence presented by \cite{Takayanagi:2012kg,Calabrese:2004eu,Ryu:2006ef} is possible to calculate the entropy S$_{A}$ in a CFT$_{2}$, holographically --- See Fig.~ \ref{3.0}. This entropy is calculated as follows:
\begin{eqnarray}
S_{A}=Min_{\Sigma_{A}}\left[\frac{Area(\Sigma_{A})}{4G_{N}}\right]\label{ES} 
\end{eqnarray}
Note that $\Sigma_{A}$ represents a two-dimensional surface, that is, four-dimensional in AdS$_{3}$ satisfying $\partial\Sigma_{A}=\partial A$ since $\Sigma_{A}$ is a counterpart to $A$. In addition to these considerations, we have that (\ref{ES}) is obtained for all surfaces, $\Sigma_{A}$, which is a minimal surface. The equation (\ref{ES}) is applied to any static configuration. Besides, as we know, the minimum surface surrounding the area is well defined in the static case, thus making it possible to consider Euclidean AdS space equivalently.
It has an interesting feature that is a striking similarity to the Bekenstein-Hawking (BH) entropy of black holes \cite{Takayanagi:2012kg,Calabrese:2004eu,Ryu:2006ef,Ryu:2006bv,Susskind:1994sm}. Besides, as we know, the minimum surface surrounding the area is well defined in the static case, thus making it possible to consider Euclidean AdS space equivalently. Motivated by the applications of \cite{Chaturvedi:2016kbk}, we propose a scenario of entanglement entropy in Horndeski gravity where is considered an analytically the entanglement
entropy of the subsystem $A$ in the ($2+1$)-dimensional boundary field theory. In our case we consider the duality AdS$_{4}$/CFT$_{3}$, where on the gravity side, we have a planar black hole solutions in Horndeski gravity and with this solution we extract the length and the area integral to the subsystem $A$. 

The main motivation to investigate the play role of Horndeski gravity in the entanglement entropy scenario is due the recent investigations of the AdS/CFT correspondence in this gravity \cite{Santos:2020egn,Brito:2019ose,Santos:2021orr,Jiang:2017imk,Baggioli:2017ojd,Liu:2018hzo,Li:2018kqp,Li:2018rgn,Feng:2015oea,Caceres:2017lbr,Hajian:2020dcq,MohammadiMozaffar:2016vpf}. Besides, beyond to the classes of boundary field theories that were studied by Ryu and Takayanagi \cite{Ryu:2006ef,Ryu:2006bv} other classes were proposed using this conjecture, such as AdS black holes with dual charge in the bulk \cite{Tonni:2010pv,Mansoori:2015sit,Caputa:2013lfa}. But in our case, we compute the entanglement entropy in AdS$_{4}$/CFT$_{3}$ within Horndeski gravity for spherically and planar black holes. For this boundary field theory at finite temperature, we address the issues to related to entanglement thermodynamics within Horndeski gravity in our setup, for more discussion about this see \cite{Blanco:2013joa,Park:2015afa,He:2014lfa,Chaturvedi:2016kbk}. For the avaible the excited states we compute the stress-energy tensor of boundary field theory in Horndeski gravity following the prescription of \cite{Santos:2021orr,Balasubramanian:1999re}.

This work is summarized as follows. In Sec.~\ref{v0} we address the issue of finding black hole solutions in Horndeski gravity. In Sec.~\ref{v1}, we present the entanglement entropy. In Sec.~\ref{v2} we compute the entanglement thermodynamics in Horndeski gravity. In section Sec.~\ref{v2}, we present our conclusions.  

\section{Black hole solutions in Horndeski gravity}\label{v0}

In this section we address the issue of finding black hole solutions in Horndeski gravity \cite{Santos:2020egn,Brito:2019ose,Santos:2021orr,Horndeski:1974wa,Santos:2020xox,Cisterna:2014nua,Bravo-Gaete:2014haa,Anabalon:2013oea}. Black holes in Horndeski's theory have been previously studied in \cite{Santos:2020egn,Brito:2019ose,Santos:2021orr,Santos:2020xox,Cisterna:2014nua,Anabalon:2013oea}. 
The Horndeski Lagrangian is given by 

\begin{eqnarray}
&&\mathcal{L}_{H}=\mathcal{L}_{2}+\mathcal{L}_{3}+\mathcal{L}_{4}+\mathcal{L}_{5},\\
&&\mathcal{L}_{2}=G_{2}(X,\phi),\\
&&\mathcal{L}_{3}=-G_{3}(X,\phi)\Box\phi,\\
&&\mathcal{L}_{4}=G_{4}(X,\phi)R+\partial_{X}G_{4}(X,\phi)\delta^{\mu\nu}_{\alpha\beta}\nabla^{\alpha}_{\mu}\phi\nabla^{\beta}_{\nu}\phi,\\
&&\mathcal{L}_{5}=G_{5}(X,\phi)G_{\mu\nu}\nabla^{\mu}\nabla^{\nu}\phi\nonumber\\
&&-\frac{1}{6}\partial_{X}G_{5}(X,\phi)\delta^{\mu\nu\rho}_{\alpha\beta\gamma}\nabla^{\alpha}_{\mu}\phi\nabla^{\beta}_{\nu}\phi\nabla^{\gamma}_{\rho}\phi,
\end{eqnarray}
where $X\equiv -\frac{1}{2}\nabla_{\mu}\phi\nabla^{\nu}\phi$. Furthermore, an interesting special truncation of this theory was presented by \cite{Charmousis:2011bf,Charmousis:2011ea,Starobinsky:2016kua,Bruneton:2012zk},where the idea is realize a constrain the coefficients $G_{k}(X,\phi)$. Through this truncation and considering the non-minimal kinetic coupling, we have 

\begin{eqnarray}
&&I[g_{\mu\nu},\phi]=\int{\sqrt{-g}d^{4}x\mathcal{L}}.\label{EH}\\
&&\mathcal{L}=\kappa(R-2\Lambda)-\frac{1}{2}(\alpha g_{\mu\nu}-\gamma G_{\mu\nu})\nabla^{\mu}\phi\nabla^{\nu}\phi\nonumber
\end{eqnarray}
Here the action (\ref{EH}) $\kappa=(16\pi G)^{-1}$. Such action have a non-minimal scalar-tensor coupling and we can define a new field $\phi^{'}\equiv\psi$. The field has a dimension of $(mass)^{2}$ controlled by the parameters $\alpha$ and $\gamma$ where $\alpha$ is dimensionless and $\gamma$ has a dimension of $(mass)^{-2}$. The equations of motion are:
\begin{equation}
G_{\mu\nu}+\Lambda g_{\mu\nu}=\frac{1}{2\kappa}T_{\mu\nu},\label{EH1}
\end{equation}
where $T_{\mu\nu}=\alpha T^{(1)}_{\mu\nu}+\gamma T^{(2)}_{\mu\nu}$. The energy-momentum tensors $T^{(1)}_{\mu\nu}$ and $T^{(2)}_{\mu\nu}$ take the following form
%\begin{equation}\begin{array}{rclrcl}
\begin{eqnarray}
&&T^{(1)}_{\mu\nu}=\nabla_{\mu}\phi\nabla_{\nu}\phi-\frac{1}{2}g_{\mu\nu}\nabla_{\lambda}\phi\nabla^{\lambda}\phi\nonumber\\
&&T^{(2)}_{\mu\nu}=\frac{1}{2}\nabla_{\mu}\phi\nabla_{\nu}\phi R-2\nabla_{\lambda}\phi\nabla_{(\mu}\phi R^{\lambda}_{\nu)}-\nabla^{\lambda}\phi\nabla^{\rho}\phi R_{\mu\lambda\nu\rho}\nonumber\\
&&-g_{\mu\nu}\left[-\frac{1}{2}(\nabla^{\lambda}\nabla^{\rho}\phi)(\nabla_{\lambda}\nabla_{\rho}\phi)+\frac{1}{2}(\Box\phi)^{2}-(\nabla_{\lambda}\phi\nabla_{\rho}\phi)R^{\lambda\rho}\right]\nonumber\\
&&-(\nabla_{\mu}\nabla^{\lambda}\phi)(\nabla_{\nu}\nabla_{\lambda}\phi)+(\nabla_{\mu}\nabla_{\nu}\phi)\Box\phi+\frac{1}{2}G_{\mu\nu}(\nabla\phi)^{2}.\label{g}
\end{eqnarray}
%\end{array}\end{equation}
And the scalar field equation is given by 
\begin{equation}
\nabla_{\mu}[(\alpha g^{\mu\nu}-\gamma G^{\mu\nu})\nabla_{\nu}\phi]=0.\label{EH2}
\end{equation}
In our case for Einstein-Horndeski gravity, we consider the following {\sl Ansatz} for a four-dimensional black hole of the form
\begin{equation}
ds^{2}=\mathcal{R}^{2}\left(-r^{2}f(r)dt^{2}+r^{2}(dx^{2}+dy^{2})+\frac{dr^{2}}{r^{2}f(r)}\right).\label{me}
\end{equation}
Now, following the results of \cite{Santos:2020egn,Brito:2019ose,Santos:2021orr,Santos:2020xox,Cisterna:2014nua,Anabalon:2013oea}, we can find the black hole solution through the imposing of the radial component with the conserved current that vanishes identically without restricting the radial dependence of the scalar field 	\cite{Hui:2012qt,Bravo-Gaete:2013dca,Babichev:2013cya}:
\begin{equation}
\alpha g_{rr}-\gamma G_{rr}=0\label{0a}.
\end{equation}
Taking $\phi'(r)\equiv \psi(r)$ and we can easily note that this condition annihilates $\psi^2(r)$ regardless of its behavior at the horizon. Now, using the equation (\ref{0a}) the metric function $f(r)$ can be found as following:
\begin{eqnarray}
f(r)&=&\frac{\alpha \mathcal{R}^{2}}{3\gamma}-\left(\frac{r_{h}}{r}\right)^{3},\label{scalar.1}\\
\psi^{2}(r)&=&-\frac{2\mathcal{R}^{2}\kappa(\alpha+\gamma\Lambda)}{\alpha\gamma r^{2}f(r)},\label{scalar.2}
\end{eqnarray}
The equation of motion (\ref{EH1})  are satisfied by these equations. The solution (\ref{scalar.1}) corresponds to black hole solution for asymptotically AdS$_4$ spacetime \cite{Anabalon:2013oea}. Beyond, through $\Lambda=-3/\mathcal{R}^{2}$ and considering the analyses of \cite{Santos:2020egn,Brito:2019ose,Santos:2021orr} in our prescription, we can write the black hole solution 
\begin{eqnarray}
f(r)&=&-\frac{\alpha}{\gamma\Lambda}-\left(\frac{r_{h}}{r}\right)^{3},\label{il}\\
\psi^{2}(r)&=&\frac{6\kappa(\alpha+\gamma\Lambda)}{\alpha\gamma\Lambda r^{2}f(r)},\label{3w}
\end{eqnarray}
we can note that the parameters are defined in the range $-\infty<\alpha/\gamma\Lambda\leq-1$, with $\alpha,\gamma<0$, or $-1\leq\alpha/\gamma\Lambda<0$, with $\alpha,\gamma>0$. Furthermore, we have that the surface located at $ r=r_{h}$ is infinitely shifted to red in relation to an asymptotic observer. On the other hand, we can see by looking at the equation (\ref{3w}), which $(\alpha+\gamma\Lambda)>0$, indicates ghost freedom. Thus, stability requires that $(\alpha+\gamma\Lambda)$ is not negative, this leads to an interval of the form $-\infty<\gamma\leq\alpha/(-\Lambda)$. Thus, following these transformation in the metric (\ref{me}) with solution (\ref{il}), we can write: 
\begin{eqnarray}
&&f(r)\to-\frac{\alpha}{\gamma\Lambda}{f}(r),r_h^3\to-\frac{\alpha}{\gamma\Lambda}\;r_h^{3},\nonumber\\
&&\mathcal{R}\to\left(-\frac{{\alpha}}{\gamma\Lambda}\right)^{1/2}\mathcal{R},t\to-\frac{\gamma\Lambda}{\alpha}t,\nonumber\\
&&x\to\left(-\frac{{\gamma\Lambda}}{\alpha}\right)^{1/2}x,y\to\left(-\frac{{\gamma\Lambda}}{\alpha}\right)^{1/2}y,
\end{eqnarray}
in order to put the black hole solution in the standard form
\begin{eqnarray}
{f}(r)&=&1-\left(\frac{r_{h}}{r}\right)^{3},\label{il2}\\
\psi^{2}(r)&=&-\frac{6\kappa(\alpha+\gamma\Lambda)}{\alpha^2 r^{2}{f}(r)}.\label{3w2}
\end{eqnarray}
In the limit $\psi^2(r\to\infty)=0$ into the action (\ref{EH}), provide that this is a genuine vacuum solution. The equations (\ref{il2})-(\ref{3w2}), provide that the black hole geometry is regular everywhere (except at the central singularity), the scalar field derivative $\psi(r)$ diverges at horizon \cite{Anabalon:2013oea,Feng:2015oea,Babichev:2013cya}, but the scalar field does not explode at horizon since it approaches to a constant near the horizon as:
\begin{eqnarray}
\phi^{2}(r)\sim\left((2\Lambda \mathcal{R}^2(\alpha+\gamma\Lambda)/\alpha^{2}r^{2}_{h}f^{'}(r_{h}))(r-r_{h})\right)+const.
\end{eqnarray}  
Thus, we agree with the no-hair theorem, such discussions were presented by \cite{Babichev:2013cya}. An interesting characterize is that the scalar field equation (\ref{3w2}) is a real function outside the horizon for $r>r_{h}$, namely, $f(r>r_{h})>0$, and the scalar field is real in the interval $-1<\alpha/\gamma\Lambda<0$, with $\alpha, \gamma>0$. Besides of this analyzes, we have that in the infinity the scalar field itself diverges as $\phi(r)\sim \ln{r}$, but not its derivatives $\psi$ that are the ones present in the action (\ref{EH}), which are finite at asymptotic infinity \cite{Babichev:2013cya}.

In fact, we have that the appearance of a black hole, which has a flat horizon $\Re^{2}$, leads us to a physics in IR that corresponds to placing the invariant theory of scale at a finite temperature. So, we have that the temperature of the black hole is given by
\begin{eqnarray}
T(r_{h})=\frac{f^{'}(r=r_{h})}{4\pi}=\frac{3}{4\pi r_{h}}\label{10}
\end{eqnarray}

\section{Holographic Entanglement Entropy in Horndeski Gravity}\label{v1}

\begin{figure}[!ht]
 \centerline{\includegraphics[scale=0.5]{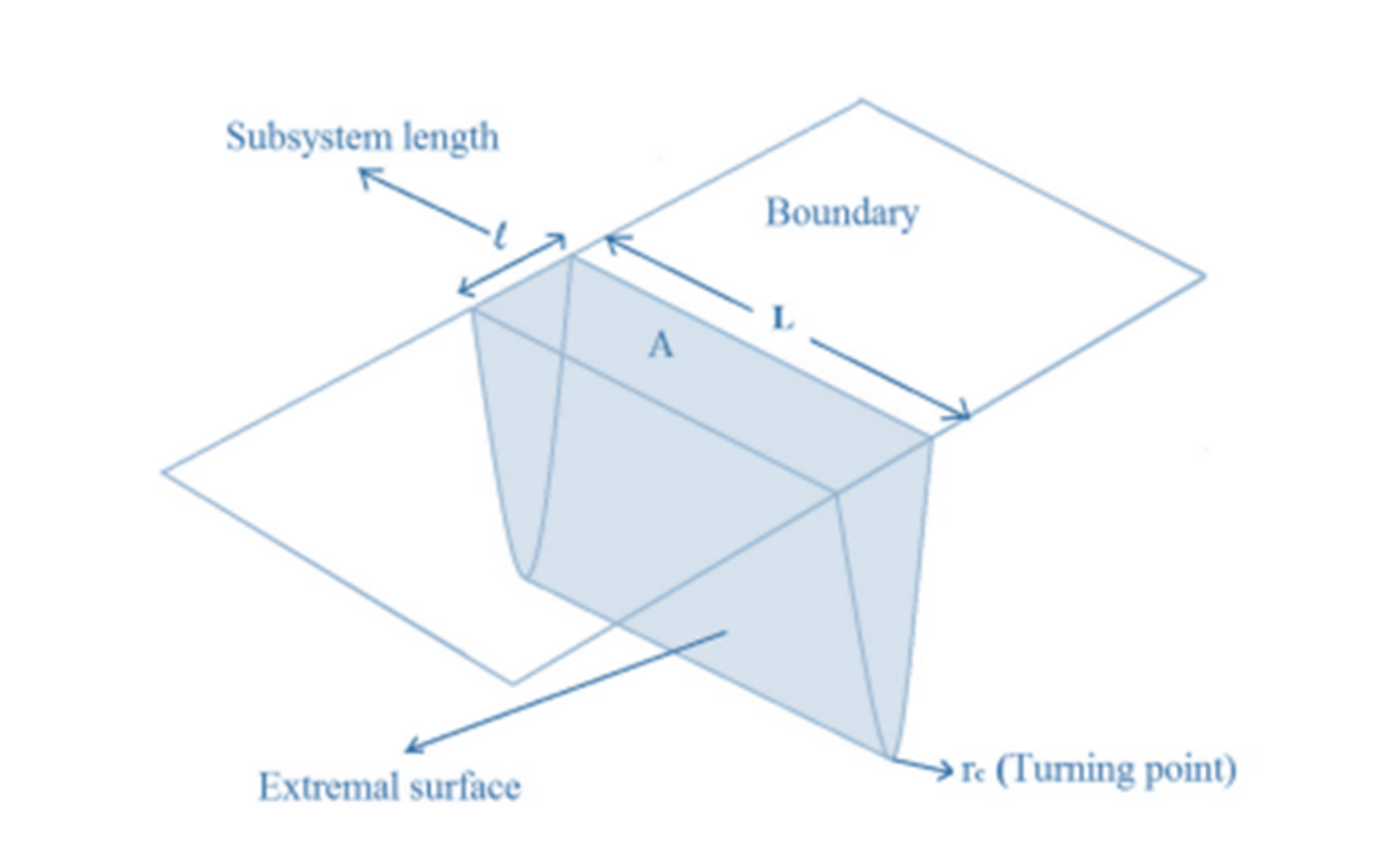}}
\caption{The figure show the schematic of a extremal surface where $l$ is the length of the subsystem $A$, which is anchored on the subsystem living on the boundary.}\label{SCHEM}
\label{planohwkhz}
\end{figure}

In this section, we present the computations of the entanglement entropy to a subsystem $A$ in Horndeski gravity following the procedures of \cite{Takayanagi:2012kg,Calabrese:2004eu,Ryu:2006ef,Ryu:2006bv,Susskind:1994sm}. For this, considering the metric (\ref{me}) the three-dimensional CFT lives in the space measured by $t$ and $x$. Thus, we can choose the subsystem $A$ (which are the two charges) to be the length $l$ interval $x\epsilon[-l/2,l/2]$, $y\epsilon[L/2,L/2]$ in the infinitely long total space $-\infty<x<\infty$, see Fig.$\sim$\ref{SCHEM}. 

Now, we will provide the entanglement entropy for the Horndeski gravity following the steps by steps of \cite{Caceres:2017lbr,Hajian:2020dcq} and motivated by recent studies of \cite{Santos:2021orr}, we have to the action (\ref{EH}) that the action with boundary counterterms is given by:
\begin{eqnarray}
&&I_{E}=I_{bulk}=-\int{\sqrt{g}d^{4}x\mathcal{L}}-2\kappa\int{d^{3}x\sqrt{\bar{\gamma}}\mathcal{L}_{b}}+2\kappa\int{d^{3}x\sqrt{\bar{\gamma}}\mathcal{L}_{ct}},\label{T1}\\
&&\mathcal{L}=\kappa(R-2\Lambda)+\frac{\gamma}{2}G_{\mu\nu}\nabla^{\mu}\phi\nabla^{\nu}\phi\label{T2}\\
&&\mathcal{L}_{b}=K^{({\bar{\gamma}})}-\Sigma^{(\bar{\gamma})}+\frac{\gamma}{4}\left(\nabla_{\mu}\phi\nabla_{\nu}\phi\, n^{\mu}n^{\nu}-(\nabla\phi)^{2}\right)K^{(\bar{\gamma})}+\frac{\gamma}{4}\nabla^{\mu}\phi\nabla^{\nu}\phi K^{(\bar{\gamma})}_{\mu\nu}\label{T3}\\
&&{\cal L}_{ct}=c_{0}+c_{1}R+c_{2}R^{ij}R_{ij}+c_{3}R^{2}+b_{1}(\partial_{i}\phi\partial^{i}\phi)^{2}\label{T4}
\end{eqnarray}
 $\mathcal{L}_{b}$ corresponds to the Gibbons-Hawking $\gamma$-dependent terms associated with the Horndeski gravity, where $n^{\mu}$ is an outward pointing unit normal vector to the boundary, $K^{(\bar{\gamma})}=\bar{\gamma}^{\mu\nu}K^{({\bar{\gamma}})}_{\mu\nu}$ is the trace of the extrinsic curvature and $\bar{\gamma}_{\mu\nu}$ is the induced metric on the boundary $r\to\infty$. The Lagragian ${\cal L}_{ct}$ is related to the boundary counterterms, they do not affect the bulk dynamics and will be neglect \cite{Balasubramanian:1999re}. Thus, the induced metric to (\ref{me}) is written as
\begin{eqnarray}
ds^{2}_{ind}=\bar{\gamma}_{\mu\nu}dx^{\mu}dx^{\nu}=\mathcal{R}^{2}\left(r^{2}f(r)d\tau^{2}+r^{2}(dx^{2}+dy^{2})+\frac{dr^{2}}{r^{2}f(r)}\right),\label{T5}
\end{eqnarray} 
With the above, we can provide that the Ryu-Takayanagi formula \cite{Ryu:2006ef,Ryu:2006bv} is given by:
\begin{eqnarray}
&&S_{A}=\frac{\mathcal{A}}{4G_{N}}\label{ES01},\\
&&\mathcal{A}=\int{ds_{ind}\chi},\\
&&\chi=1-2\gamma(\bar{\gamma}^{\lambda\sigma}\nabla_{\lambda}\phi\nabla_{\sigma}\phi).
\end{eqnarray}
Here this entropy is equivalently to the cases of \cite{Feng:2015oea,Caceres:2017lbr}. For AdS$_{4}$/CFT$_{3}$, we have that the minimum surface is given by the geodesic line in AdS$_{4}$. That is, we have following the prescription of \cite{Chaturvedi:2016kbk}, we have that the area of the surface anchored on the boundary to the subsystem (A), can be expressed as: 
\begin{eqnarray}
\mathcal{A}=2\chi\mathcal{R}Lr_{c}\int^{1}_{0}{\frac{du}{u^{2}\sqrt{(1-u^{4})f(u)}}};\quad f(u)=1-\left(\frac{r_{h}u}{r_{c}}\right)^{3},\label{ES1}
\end{eqnarray}
where $u=r_{c}/r$. In our setup to holographic entanglement entropy in Horndeski gravity $r_{c}$ is a constant of integration, which represents the turning point of the extremal surface in the higher dimensional AdS4 bulk spacetime, see Fig.$\sim$\ref{SCHEM}. We can see that through the equation (\ref{ES1}) the area integral is divergent at the point $u=1$ and must be regularized introducing an infrared cutoff ($r_{b}$). From the point of view of the holographic dictionary, we have a relation between the UV cutoff of the boundary field theory ($\epsilon$) to bulk IR cutoff. Such relation is inversely related through the AdS length scale $\mathcal{R}$ and can be establish as, $r_{b}=\mathcal{R}/\epsilon$. But note that the finite part of entanglement entropy can then be used to study the high and low temperature behavior of the boundary field theory which is dual to black hole as covered in
\begin{eqnarray}
&&S^{finite}_{A}=S_{A}-S^{divergent}_{A}=\frac{\mathcal{A}^{finite}}{4G_{N}}\chi.\label{ES02}\\
&&\chi=1+\frac{12\kappa\gamma(\alpha+\gamma\Lambda)}{\alpha^{2}\mathcal{R}^{2}}.
\end{eqnarray}
We can to obtain the quantity $r_{c}$ inverting the equation of motion
\begin{eqnarray}
\frac{l}{2}=\frac{1}{r_{c}}\int^{1}_{0}{\frac{u^{2}du}{\sqrt{(1-u^{4})f(u)}}}\label{ES2} 
\end{eqnarray}
As we know, the horizon radius ($r_{h}$) is very small, we have that the black hole remains deep inside the bulk. Thus, far away from the extremal surface, namely, $r_{h}<<r_{c}$.For this limit, we can performing Taylor expand. In this case the quantity $1/\sqrt{f(u)}$ will be expanded around $r_{h}/r_{c}=0$ as:
\begin{eqnarray}
\frac{1}{\sqrt{f(u)}}=1+\frac{1}{2}\left(\frac{r_{h}u}{r_{c}}\right)^{3}+\mathcal{O}\left[\left(\frac{r_{h}u}{r_{c}}\right)^{3}\right]^{4} 
\end{eqnarray}
Now, replacing this expansion in the equation (\ref{ES2}), we can wrote
\begin{eqnarray}
&&\frac{lr_{c}}{2}=\int^{1}_{0}{\frac{u^{2}du}{\sqrt{(1-u^{4})}}}+\frac{1}{2}\left(\frac{r_{h}}{r_{c}}\right)^{3}\int^{1}_{0}{\frac{u^{5}du}{\sqrt{(1-u^{4})}}}\label{ES3}\\
&&\frac{lr_{c}}{2}=\pi\left(\frac{r_{h}}{r_{c}}\right)^{3}+\frac{2\sqrt{\pi}\Gamma(3/4)}{\Gamma(1/4)}\label{ES4}
\end{eqnarray}
However, solving the equation (\ref{ES4}) in terms of $r_{h}l$, we have
\begin{eqnarray}
r_{c}=\frac{2\pi\Gamma(3/4)}{l\Gamma(1/4)}+\frac{l^{2}\Gamma^{3}(3/4)}{16\sqrt{\pi}\Gamma^{3}(1/4)}\label{ES5}
\end{eqnarray}
For the extremal area, we can write following the same steps
\begin{eqnarray}
\mathcal{A}=2\chi\mathcal{R}Lr_{c}\left(\int^{1}_{0}{\frac{u^{2}du}{\sqrt{(1-u^{4})}}}+\frac{1}{2}\left(\frac{r_{h}}{r_{c}}\right)^{3}\int^{1}_{0}{\frac{udu}{\sqrt{(1-u^{4})}}}\right)\label{ES6}
\end{eqnarray}
Through the equation (\ref{ES6}), we have that first integral is same of a pure AdS, which is divergent. For this reason, we need regularize her, introducing an UV cutoff $1/r_{b}$ and add a counter term ($-2\mathcal{R}Lr_{b}$), after this considerations we provide the finite part of the extremal area as
\begin{eqnarray}
&&\mathcal{A}^{finite}=2\chi\mathcal{R}Lr_{c}\left(\int^{1}_{0}{\frac{u^{2}du}{\sqrt{(1-u^{4})}}}-2\mathcal{R}Lr_{b}+\frac{1}{2}\left(\frac{r_{h}}{r_{c}}\right)^{3}\int^{1}_{0}{\frac{udu}{\sqrt{(1-u^{4})}}}\right)\label{ES7}\\
&&\mathcal{A}^{finite}=\chi\mathcal{R}Lr_{c}\left(\sqrt{\pi}\frac{\Gamma(-1/4)}{\Gamma(1/4)}+\pi\left(\frac{r_{h}}{r_{c}}\right)^{3}\right)\label{ES8}
\end{eqnarray}
Combining the equation (\ref{ES8}) with (\ref{ES5}) the entanglement entropy is given by:
\begin{eqnarray}
S^{finite}_{A}=\frac{\chi\mathcal{R}L}{4lG_{N}}\left(-\frac{4\pi\Gamma^{2}(3/4)}{\Gamma^{2}(1/4)}+\frac{l^{3}r^{3}_{h}\Gamma^{2}(1/4)}{8\Gamma^{2}(3/4)}\right)\label{ES90}
\end{eqnarray}
For case of extremal black holes we consider $r^{3}_{h}=\mathcal{M}_{ext}/4$ where $\mathcal{M}_{ext}$ is the mass of the extremal black hole, we have
\begin{eqnarray}
&&S^{finite}_{A}=[S^{AdS}_{A}+k\mathcal{M}_{ext}l^{2}L]\chi,\quad k=\frac{\mathcal{R}L}{32G_{N}}\frac{\Gamma^{2}(1/4)}{\Gamma^{2}(3/4)}\label{ES10}\\
&&S^{AdS}_{A}=-\frac{4\pi\mathcal{R}L}{4lG_{N}}\frac{\Gamma^{2}(3/4)}{\Gamma^{2}(1/4)}\label{ES11}
\end{eqnarray}
The results show in the equations (\ref{ES10}) and (\ref{ES11}) are similar to the case of \cite{Chaturvedi:2016kbk}, but in our case, we have the presence of Horndeski parameters, where $S^{AdS}_{A}$ is the entanglement entropy of the subsystem (A), when the bulk theory is pure AdS as described by \cite{Fischler:2012ca}. On the other hand, if $\gamma$ is large the Horndeski contribution through the $\chi$-term in the equation (\ref{ES10}), this imply that $S^{finite}_{A}$ becomes large and the storage of information show that, we have more information about the subsystem (A). Furthermore, at the critical point  $\alpha=3\gamma/\mathcal{R}^{2}$, for more discussion see \cite{Li:2018kqp}, the entropy in equation (\ref{ES10}) reduces to the usual case of \cite{Chaturvedi:2016kbk} where sub-leading correction term in the equation (\ref{ES10}), becomes important in defining the first law like relation.

\subsection{Planar black hole}

We address the issue of finding planar black hole solutions  in Horndeski gravity. For this, we consider for Einstein-Horndeski gravity the following {\sl Ansatz}:

\begin{equation}
ds^{2}=-f(r)dt^{2}+r^{2}(dx^{2}+dy^{2})+\frac{dr^{2}}{f(r)}.\label{P}
\end{equation}
One can show that the equations (\ref{EH1}) and (\ref{EH2}) are satisfied by the following solution
\begin{eqnarray}
f(r)&=&\frac{\alpha r^{2}}{3\gamma}-\frac{r_{h}}{r},\label{P1}\\
\psi^{2}(r)&=&-\frac{2\kappa(\alpha+\gamma\Lambda)}{\alpha\gamma r^{2}}\frac{1}{f(r)}.\label{P2}
\end{eqnarray}
Following the previous procedures, we have
\begin{eqnarray}
&&\frac{l}{2}=\int^{1}_{0}{\frac{du}{u\sqrt{(1-u^{4})f(u)}}}; \quad f(u)=\frac{\alpha r^{2}_{c}}{3\gamma u^{2}}-\frac{r_{h}u}{r_{c}},\label{P3}\\
&&\mathcal{A}=2\chi Lr_{c}\int^{1}_{0}{\frac{du}{u^{2}\sqrt{(1-u^{4})f(u)}}}\label{P4}
\end{eqnarray}
Again through the expansion of $1/\sqrt{f(u)}$, we have
\begin{eqnarray}
\frac{1}{\sqrt{f(u)}}=\sqrt{\frac{3\gamma}{\alpha}}\frac{u}{r_{c}}+\sqrt{\left(\frac{3\gamma}{\alpha}\right)^{3}}\frac{u^{4}r_{h}}{r^{4}_{c}}\label{P5}
\end{eqnarray}
Now, we can write for the equations (\ref{P3}) and (\ref{P4}) as:
\begin{eqnarray}
&&\frac{l}{2}=\frac{1}{r_{c}}\sqrt{\frac{3\gamma}{\alpha}}\frac{\sqrt{\pi}\Gamma(5/4)}{\Gamma(3/4)}+\frac{r_{h}}{4r^{4}_{c}}\sqrt{\left(\frac{3\gamma}{\alpha}\right)^{3}},\label{P6}\\
&&\mathcal{A}^{finite}\approx \chi L\sqrt{\frac{3\gamma}{\alpha}}\ln\left(\frac{l}{r_{b}}\right)-\frac{3\sqrt{3\pi}l\chi L}{4}\sqrt{\frac{\gamma}{\alpha}}\frac{\Gamma(5/4)}{\Gamma(1/4)}\label{P7}
\end{eqnarray}
Thus, the entanglement entropy to the planar black is
\begin{eqnarray}
S^{finite}_{A}=\frac{\mathcal{A}^{finite}}{4G_{N}}=\frac{\chi L}{4G_{N}}\sqrt{\frac{3\gamma}{\alpha}}\ln\left(\frac{l}{r_{b}}\right)-\frac{3\sqrt{\pi}l\chi L}{16G_{N}}\sqrt{\frac{3\gamma}{\alpha}}\frac{\Gamma(5/4)}{\Gamma(1/4)}\label{P7}
\end{eqnarray}
When we compare the equation (\ref{P7}) with the equation (\ref{ES10}), we have that entanglement entropy to the subsystem (A) have a logarithmic term \cite{Calabrese:2004eu,Holzhey:1994we} and a sub-leading correction with Horndeski parameters. Besides, if $\gamma$ is small, the $S^{finite}_{A}\to 0$ for the planar black hole and all information about the subsystem (A) is destroyed, but not at the critical point $\alpha=3\gamma/\mathcal{R}^{2}$ \cite{Li:2018kqp} where we have a behavior like area law, which is corrected by a logarithmic factor. Such behavior is find fermionic systems with the presence of a finite Fermi surface \cite{Gioev:2006zz,Wolf:2006zzb}.

\section{Entanglement thermodynamics in Horndeski gravity}\label{v2}

In this section, we present the "first law of entanglement thermodynamics". For this, we need of the stress-energy tensor of boundary field theory in Horndeski gravity. Through the renormalization procedure \cite{Balasubramanian:1999re} the form of stress-energy tensor $T_{\alpha\beta}$ can be write as:
\begin{eqnarray}
&&T_{\alpha\beta}=-\frac{r^{3}}{16\pi\mathcal{R}^{3} G_{N}}\left[K^{({\bar{\gamma}})}_{\alpha\beta}-\bar{\gamma}_{\alpha\beta}(K^{({\bar{\gamma}})}-\Sigma^{(\bar{\gamma})})+\frac{\gamma}{4}H_{\alpha\beta}-\kappa T^{R}_{\alpha\beta}-\kappa T^{ct}_{\alpha\beta}\right]\label{T},\label{T6}\\
&&H_{\alpha\beta}=(\nabla_{\alpha}\phi\nabla_{\beta}\phi n^{\alpha}n^{\beta}-(\nabla\phi)^{2})(K^{({\bar{\gamma}})}_{\alpha\beta}-\bar{\gamma}_{\alpha\beta}K^{({\bar{\gamma}})})-(\nabla_{\alpha}\phi\nabla_{\beta}\phi)\bar{\gamma}_{\alpha\beta}K^{({\bar{\gamma}})}.\label{T7}
\end{eqnarray}
Here $T^{R}_{\alpha\beta}$ and $T^{ct}_{\alpha\beta}$ are possible contribution of extrinsic curvature and counter term, respectively. However, fixing the energy-momentum tensor on the boundary with $T^{R}_{\alpha\beta}=T^{ct}_{\alpha\beta}=0$, we have
\begin{eqnarray}
T_{\alpha\beta}=-\frac{r^{3}}{16\pi\mathcal{R}^{3} G_{N}}\left[K^{({\bar{\gamma}})}_{\alpha\beta}-\bar{\gamma}_{\alpha\beta}(K^{({\bar{\gamma}})}-\Sigma^{(\bar{\gamma})})+\frac{\gamma}{4}H_{\alpha\beta}\right].\label{T8}
\end{eqnarray}
However, in order to obtain the “first law of entanglement thermodynamics” in Horndeski gravity we need compute the following quantity:
\begin{eqnarray}
&&\Delta S_{A}=\frac{\Delta E_{A}}{T_{en}},\label{T9}\\
&&\Delta E_{A}=\int_{A}{dxdyT^{Temp\neq 0}_{tt}}-\int_{A}{dxdyT^{Temp=0}_{tt}}.\label{T10}
\end{eqnarray}
But one interesting physical observables is the mass written as:
\begin{eqnarray}
\mathcal{M}=\int_{A}{dxdyT^{Temp\neq 0}_{tt}}.\label{T10}
\end{eqnarray}
With
\begin{eqnarray}
&&T_{tt}=-\chi\frac{r^{3}}{16\mathcal{R}^{3}\pi G_{N}}\left[K^{({\bar{\gamma}})}_{tt}-\bar{\gamma}_{tt}(K^{({\bar{\gamma}})}-\Sigma^{(\bar{\gamma})})\right],\label{T80}\\
&&T_{tt}=\chi\frac{r^{3}}{8\pi\mathcal{R}^{4}G_{N}}.\label{T81}
\end{eqnarray}
The result presented by the equation (\ref{T81}) is the same of \cite{Balasubramanian:1999re} where the equation (\ref{T81}) corresponds to a excited state in the CFT$_{3}$ \cite{Bhattacharya:2012mi}. In this way, we can express the first law of entanglement thermodynamics as:
\begin{eqnarray}
&&\Delta E_{A}=\frac{l\chi L}{8\pi G_{N}}(\mathcal{M}-\mathcal{M}_{ext}),\label{T9}\\
&&T_{en}=\frac{\pi}{16l}\frac{\Gamma^{2}(3/4)}{\Gamma^{2}(1/4)}.\label{T10}
\end{eqnarray}
Where $T_{en}$ is the  entanglement temperature \cite{Bhattacharya:2012mi,Chaturvedi:2016kbk}. Such result are in perfect agreement with \cite{Blanco:2013joa}. But here this result and  also as shown in  \cite{Bhattacharya:2012mi}, it differs from our results, because we consider a canonical ensemble where the ground state of the boundary field theory is dual to the extremal AdS$_{4}$ black hole in the bulk.
\subsection{Planar black hole}
Now, for the planar black hole, following the step by step as done to the spherically case we compute the increased amount of energy for such black hole. Thus, through the equation (\ref{T8}), we can show that: 
\begin{eqnarray}
T_{tt}=\frac{\chi}{16\pi\mathcal{R}G_{N}}\sqrt{\frac{3\gamma}{\alpha}}\left(\frac{4}{r}-\frac{1}{R}\right).\label{T11}
\end{eqnarray}
The equation (\ref{T11}), provide that we can must be not need to make any assumptions in the infrared region $r\to\infty$ for this excited state in the CFT$_{3}$. Thus, objects such as black branes or stars in the infrared region can be find. The first has a horizon and is a thermal state, but the second does not have any horizon, however, is dual to a zero temperature state \cite{Bhattacharya:2012mi}. Furthermore, the increased amount of energy in the subsystem (A) can be written as
\begin{eqnarray}
&&\Delta E_{A}=\frac{\chi}{16\pi\mathcal{R}G_{N}}\sqrt{\frac{\alpha}{3\gamma}}\int_{A}{dydr\frac{4}{r}}-\frac{\chi}{16\pi\mathcal{R}^{2}G_{N}}\sqrt{\frac{\alpha}{3\gamma}}\int_{A}{dxdy},\\
&&\Delta E_{A}=\frac{\chi L}{4\pi\mathcal{R}G_{N}}\sqrt{\frac{3\gamma}{\alpha}}\ln\left(\frac{l}{r_{b}}\right)-\frac{l\chi L}{16\pi\mathcal{R}^{2}G_{N}}\sqrt{\frac{3\gamma}{\alpha}},\label{T12}\\
&&T_{en}=\frac{1}{3\sqrt{\pi}l}\frac{\Gamma(1/4)}{\Gamma(5/4)}.\label{T13}
\end{eqnarray}
where the entropy in equation (\ref{T12}) to the planar black hole is the area of the boundary in the region $r\to\infty$. In summary, this result is very similar to the equation (\ref{T9}). On the other hand, when we compare the equation (\ref{T12}) with the equation (\ref{P7}), we see that are very similar. This convergence of results can be done through the renormalization procedure, which removes the logarithmic correction.  

\section{Conclusion}\label{v3}

We show in four-dimensions that the study of the entanglement entropy in Horndeski gravity for planar and spherically topologies for a strip like region denoted by the subsystem (A), which is Boundary Conformal Field Theory dual to bulk black holes in an AdS$_{4}$/CFT$_{3}$ scenario, provided interesting results, respectively. For this two topologies, we find two interesting aspects of the holographic entanglement entropy. The first behavior of the entanglement entropy with the Horndeski parameters of the AdS black hole can be becomes pure AdS, if kinematic coupling $\gamma$ is very small, that is, $\gamma\to 0$. Thus, such result show that there is similar behavior of study in the field of AdS/CFT correspondence within Horndeski gravity, when if explore the black hole thermodynamics of the black holes, as for example in the studies of \cite{Santos:2021orr,Feng:2015oea}. For $\gamma=0$ we recover the usual entropy of the AdS space. 

The second behavior for the entanglement entropy is very interesting, because the values of parameters of Horndeski gravity impose limitations in storage of information to the planar black holes, if the entanglement entropy of him becomes null, that is, $S^{finite}_{A}\to 0$ when $\gamma\to 0$ the storage information are completely destroyed, but in the critical points \cite{Li:2018kqp} where $\alpha=3\gamma/\mathcal{R}^{2}$, we have that the entanglement entropy's described by equation (\ref{P7}) preserve the storage information of the subsystem (A), because are not constraint by the parameters at this point.

Finally, for the “first law of entanglement thermodynamics” in Horndeski gravity, we show that for spherically case, it has extremal black holes solution. The first topology of extremal black hole the large implies a large horizon radius. For the second topology, we have that the increasing of energy $\Delta E_{A}$ is constraint by the Horndeski parameters. This fact, agree with Ryu-Takayanagi formula as shown in (\ref{P7}). This first law of entanglement thermodynamics in Horndeski gravity in fact agree with the result of \cite{Santos:2021orr,Feng:2015oea,Caceres:2017lbr,Hajian:2020dcq}.

\section*{Acknowledgment}

Fabiano F. Santos would like to thank CNPq and CAPES for partial financial support. I would like to thank Mohammad Hassan Vahidinia for valuable comments and discussions.


\newpage    
\begin{thebibliography}{99}

%\cite{Takayanagi:2012kg}
\bibitem{Takayanagi:2012kg}
T.~Takayanagi,
{\it Entanglement Entropy from a Holographic Viewpoint},
Class. Quant. Grav. \textbf{29} (2012), 153001,
[arXiv:1204.2450 [gr-qc]].
%178 citations counted in INSPIRE as of 08 Jul 2020

%\cite{Bhattacharya:2019zkb}
\bibitem{Bhattacharya:2019zkb}
A.~Bhattacharya, K.~T.~Grosvenor and S.~Roy,
{\it Entanglement Entropy and Subregion Complexity in Thermal Perturbations around Pure-AdS Spacetime},
Phys. Rev. D \textbf{100}, no.12, 126004 (2019),
arXiv:1905.02220 [hep-th]].
%26 citations counted in INSPIRE as of 11 Jan 2022

%\cite{Bhattacharya:2017gzt}
\bibitem{Bhattacharya:2017gzt}
A.~Bhattacharya and S.~Roy,
{\it Holographic entanglement entropy and entanglement thermodynamics of \textquoteleft{}black\textquoteright{} non-susy D3 brane},
Phys. Lett. B \textbf{781}, 232-237 (2018),
[arXiv:1712.03740 [hep-th]].
%7 citations counted in INSPIRE as of 11 Jan 2022

%\cite{Bhattacharya:2012mi}
\bibitem{Bhattacharya:2012mi}
J.~Bhattacharya, M.~Nozaki, T.~Takayanagi and T.~Ugajin,
{\it Thermodynamical Property of Entanglement Entropy for Excited States},
Phys. Rev. Lett. \textbf{110}, no.9, 091602 (2013),
[arXiv:1212.1164 [hep-th]].
%205 citations counted in INSPIRE as of 07 Jan 2022


%\cite{Calabrese:2004eu}
\bibitem{Calabrese:2004eu}
P.~Calabrese and J.~L.~Cardy,
{\it Entanglement entropy and quantum field theory},
J. Stat. Mech. \textbf{0406} (2004), P06002,
[arXiv:hep-th/0405152 [hep-th]].
%1190 citations counted in INSPIRE as of 08 Jul 2020

%\cite{Holzhey:1994we}
\bibitem{Holzhey:1994we}
C.~Holzhey, F.~Larsen and F.~Wilczek,
{\it Geometric and renormalized entropy in conformal field theory},
Nucl. Phys. B \textbf{424}, 443-467 (1994),
[arXiv:hep-th/9403108 [hep-th]].
%1031 citations counted in INSPIRE as of 05 Jan 2022

%\cite{Ryu:2006ef}
\bibitem{Ryu:2006ef}
S.~Ryu and T.~Takayanagi,
{\it Aspects of Holographic Entanglement Entropy},
JHEP \textbf{08} (2006), 045,
[arXiv:hep-th/0605073 [hep-th]].
%1287 citations counted in INSPIRE as of 08 Jul 2020

%\cite{Ryu:2006bv}
\bibitem{Ryu:2006bv}
S.~Ryu and T.~Takayanagi,
{\it Holographic derivation of entanglement entropy from AdS/CFT},
Phys. Rev. Lett. \textbf{96} (2006), 181602,
[arXiv:hep-th/0603001 [hep-th]].
%2373 citations counted in INSPIRE as of 15 Aug 2020

%\cite{Susskind:1994sm}
\bibitem{Susskind:1994sm}
L.~Susskind and J.~Uglum,
{\it Black hole entropy in canonical quantum gravity and superstring theory},
Phys. Rev. D \textbf{50} (1994), 2700-2711,
[arXiv:hep-th/9401070 [hep-th]].
%586 citations counted in INSPIRE as of 16 Jul 2020 

%\cite{Chaturvedi:2016kbk}
\bibitem{Chaturvedi:2016kbk}
P.~Chaturvedi, V.~Malvimat and G.~Sengupta,
{\it Entanglement thermodynamics for charged black holes},
Phys. Rev. D \textbf{94}, no.6, 066004 (2016),
[arXiv:1601.00303 [hep-th]].
%25 citations counted in INSPIRE as of 30 Dec 2021

%\cite{Tonni:2010pv}
\bibitem{Tonni:2010pv}
E.~Tonni,
{\it Holographic entanglement entropy: near horizon geometry and disconnected regions},
JHEP \textbf{05} (2011), 004,
[arXiv:1011.0166 [hep-th]].
%62 citations counted in INSPIRE as of 15 Aug 2020

%\cite{Mansoori:2015sit}
\bibitem{Mansoori:2015sit}
S.~A.~H.~Mansoori, B.~Mirza, M.~D.~Darareh and S.~Janbaz,
{\it Entanglement Thermodynamics of the Generalized Charged BTZ Black Hole},
Int. J. Mod. Phys. A \textbf{31} (2016) no.12, 1650067,
[arXiv:1512.00096 [gr-qc]].
%8 citations counted in INSPIRE as of 15 Aug 2020

%\cite{Caputa:2013lfa}
\bibitem{Caputa:2013lfa}
P.~Caputa, V.~Jejjala and H.~Soltanpanahi,
{\it Entanglement entropy of extremal BTZ black holes},
Phys. Rev. D \textbf{89} (2014) no.4, 046006,
[arXiv:1309.7852 [hep-th]].
%22 citations counted in INSPIRE as of 15 Aug 2020

%\cite{Blanco:2013joa}
\bibitem{Blanco:2013joa}
D.~D.~Blanco, H.~Casini, L.~Y.~Hung and R.~C.~Myers,
{\it Relative Entropy and Holography},
JHEP \textbf{08} (2013), 060,
[arXiv:1305.3182 [hep-th]].
%263 citations counted in INSPIRE as of 15 Aug 2020

%\cite{Susskind:2021esx}
\bibitem{Susskind:2021esx}
L.~Susskind,
{\it Entanglement and Chaos in De Sitter Holography: An SYK Example},
[arXiv:2109.14104 [hep-th]].
%9 citations counted in INSPIRE as of 09 Feb 2022

%\cite{Kay:2016kbi}
\bibitem{Kay:2016kbi}
B.~S.~Kay,
{\it Entanglement entropy and algebraic holography},
[arXiv:1605.07872 [hep-th]].
%0 citations counted in INSPIRE as of 09 Feb 2022

%\cite{Park:2015afa}
\bibitem{Park:2015afa}
C.~Park,
{\it Holographic entanglement entropy in the nonconformal medium},
Phys. Rev. D \textbf{91} (2015) no.12, 126003,
[arXiv:1501.02908 [hep-th]].
%19 citations counted in INSPIRE as of 15 Aug 2020

%\cite{He:2014lfa}
\bibitem{He:2014lfa}
S.~He, J.~R.~Sun and H.~Q.~Zhang,
{\it On Holographic Entanglement Entropy with Second Order Excitations},
Nucl. Phys. B \textbf{928} (2018), 160-181,
[arXiv:1411.6213 [hep-th]].
%17 citations counted in INSPIRE as of 15 Aug 2020


%\cite{Santos:2020egn}
\bibitem{Santos:2020egn}
F.~F.~Santos,
{\it Aplica\c{c}\~oes do Setor John da Gravidade de Horndeski nos Cen\'arios de Brana Negra e Rela\c{c}\~ao de viscosidade/entropia, Mundo Brana e Cosmologia (In Portuguese)},
[arXiv:2006.06550 [hep-th]].
%1 citations counted in INSPIRE as of 03 Jan 2022

%\cite{Brito:2019ose}
\bibitem{Brito:2019ose}
F.~A.~Brito and F.~F.~Santos,
{\it Black brane in asymptotically Lifshitz spacetime and viscosity/entropy ratios in Horndeski gravity},
EPL \textbf{129}, no.5, 50003 (2020),
[arXiv:1901.06770 [hep-th]].
%11 citations counted in INSPIRE as of 03 Jan 2022

%\cite{Santos:2021orr}
\bibitem{Santos:2021orr}
F.~F.~Santos, E.~F.~Capossoli and H.~Boschi-Filho,
{\it AdS/BCFT correspondence and BTZ black hole thermodynamics within Horndeski gravity},
Phys. Rev. D \textbf{104}, no.6, 066014 (2021),
[arXiv:2105.03802 [hep-th]].
%1 citations counted in INSPIRE as of 03 Jan 2022


\bibitem{Jiang:2017imk} 
  W.~J.~Jiang, H.~S.~Liu, H.~Lu and C.~N.~Pope,
  {\it DC Conductivities with Momentum Dissipation in Horndeski Theories},
  JHEP {\bf 1707}, 084 (2017),
  [arXiv:1703.00922 [hep-th]].
  %%CITATION = doi:10.1007/JHEP07(2017)084;%%
  %16 citations counted in INSPIRE as of 22 Jan 2019
	
	\bibitem{Baggioli:2017ojd} 
  M.~Baggioli and W.~J.~Li,
  {\it Diffusivities bounds and chaos in holographic Horndeski theories},
  JHEP {\bf 1707}, 055 (2017),
  [arXiv:1705.01766 [hep-th]].
  %%CITATION = doi:10.1007/JHEP07(2017)055;%%
  %32 citations counted in INSPIRE as of 22 Jan 2019

\bibitem{Liu:2018hzo} 
H.~S.~Liu,
{\it Violation of Thermal Conductivity Bound in Horndeski Theory},
Phys.\ Rev.\ D {\bf 98}, no. 6, 061902 (2018),
[arXiv:1804.06502 [hep-th]].
%%CITATION = doi:10.1103/PhysRevD.98.061902;%%
%2 citations counted in INSPIRE as of 20 Jan 2019

\bibitem{Li:2018kqp} 
  Y.~Z.~Li and H.~Lu,
  {\it $a$-theorem for Horndeski gravity at the critical point},
  Phys.\ Rev.\ D {\bf 97}, no. 12, 126008 (2018),
  [arXiv:1803.08088 [hep-th]].
  %%CITATION = doi:10.1103/PhysRevD.97.126008;%%
  %6 citations counted in INSPIRE as of 22 Jan 2019

 \bibitem{Li:2018rgn} 
 Y.~Z.~Li, H.~Lu and H.~Y.~Zhang,
 {\it Scale Invariance vs. Conformal Invariance: Holographic Two-Point Functions in Horndeski Gravity},
 arXiv:1812.05123 [hep-th].
 %%CITATION = ARXIV:1812.05123;%%
 %1 citations counted in INSPIRE as of 22 Jan 2019 
	
	\bibitem{Feng:2015oea} 
  X.~H.~Feng, H.~S.~Liu, H.~Lü and C.~N.~Pope,
  {\it Black Hole Entropy and Viscosity Bound in Horndeski Gravity},
  JHEP {\bf 1511}, 176 (2015),
  [arXiv:1509.07142 [hep-th]].
  %%CITATION = doi:10.1007/JHEP11(2015)176;%%
  %42 citations counted in INSPIRE as of 31 Dec 2018
  
  %\cite{Caceres:2017lbr}
\bibitem{Caceres:2017lbr}
E.~Caceres, R.~Mohan and P.~H.~Nguyen,
{\it On holographic entanglement entropy of Horndeski black holes},
JHEP \textbf{10}, 145 (2017)
doi:10.1007/JHEP10(2017)145
[arXiv:1707.06322 [hep-th]].
%16 citations counted in INSPIRE as of 10 Jan 2022

%\cite{Hajian:2020dcq}
\bibitem{Hajian:2020dcq}
K.~Hajian, S.~Liberati, M.~M.~Sheikh-Jabbari and M.~H.~Vahidinia,
{\it On Black Hole Temperature in Horndeski Gravity},
Phys. Lett. B \textbf{812}, 136002 (2021)
doi:10.1016/j.physletb.2020.136002
[arXiv:2005.12985 [gr-qc]].
%13 citations counted in INSPIRE as of 10 Jan 2022

%\cite{MohammadiMozaffar:2016vpf}
\bibitem{MohammadiMozaffar:2016vpf}
M.~R.~Mohammadi Mozaffar, A.~Mollabashi, M.~M.~Sheikh-Jabbari and M.~H.~Vahidinia,
{\it Holographic Entanglement Entropy, Field Redefinition Invariance and Higher Derivative Gravity Theories},
Phys. Rev. D \textbf{94}, no.4, 046002 (2016),
[arXiv:1603.05713 [hep-th]].
%22 citations counted in INSPIRE as of 10 Jan 2022
  
  %\cite{Balasubramanian:1999re}
\bibitem{Balasubramanian:1999re} 
 V.~Balasubramanian and P.~Kraus,
{\it A Stress tensor for Anti-de Sitter gravity},
Commun.\ Math.\ Phys.\  {\bf 208}, 413 (1999),
[hep-th/9902121].
%CITATION = doi:10.1007/s002200050764;%%
%1380 citations counted in INSPIRE as of 28 May 2019

%%%%%%%%%%%%%%%%%%%%%%%%%%%%%%%%%%%%%%%%%%%%%%%%%%%%%%%%%%%%%%%%%%%%%%%%%%%%%%%%%%%%%%%%%%%%%%%%%%%%%%%%%%%%%%%%%%%%

\bibitem{Horndeski:1974wa} 
  G.~W.~Horndeski,
  {\it Second-order scalar-tensor field equations in a four-dimensional space},
  Int.\ J.\ Theor.\ Phys.\  {\bf 10}, 363 (1974).
  %%CITATION = doi:10.1007/BF01807638;%%
  %1009 citations counted in INSPIRE as of 05 Jan 2019
  
  %\cite{Santos:2020xox}
\bibitem{Santos:2020xox}
F.~F.~Santos,
{\it Rotating black hole with a probe string in Horndeski Gravity},
Eur. Phys. J. Plus \textbf{135}, no.10, 810 (2020),
[arXiv:2005.10983 [hep-th]].
%6 citations counted in INSPIRE as of 03 Jan 2022
  
  	\bibitem{Cisterna:2014nua} 
  A.~Cisterna and C.~Erices,
  {\it Asymptotically locally AdS and flat black holes in the presence of an electric field in the Horndeski scenario},
  Phys.\ Rev.\ D {\bf 89}, 084038 (2014),
  [arXiv:gr-qc/1401.4479].
  %%CITATION = doi:10.1103/PhysRevD.89.084038;%%
  %73 citations counted in INSPIRE as of 31 Dec 2018
	
	%\cite{Bravo-Gaete:2014haa}
\bibitem{Bravo-Gaete:2014haa}
M.~Bravo-Gaete and M.~Hassaine,
{\it Thermodynamics of a BTZ black hole solution with an Horndeski source},
Phys. Rev. D \textbf{90}, no.2, 024008 (2014),
[arXiv:1405.4935 [hep-th]].
%38 citations counted in INSPIRE as of 11 Jan 2022
	
  \bibitem{Anabalon:2013oea} 
  A.~Anabalon, A.~Cisterna and J.~Oliva,
  {\it Asymptotically locally AdS and flat black holes in Horndeski theory},
  Phys.\ Rev.\ D {\bf 89}, 084050 (2014),
  [arXiv:gr-qc/1312.3597 [gr-qc]].
  %%CITATION = doi:10.1103/PhysRevD.89.084050;%%
  %119 citations counted in INSPIRE as of 31 Dec 2018

%\cite{Charmousis:2011bf}
\bibitem{Charmousis:2011bf} 
  C.~Charmousis, E.~J.~Copeland, A.~Padilla and P.~M.~Saffin,
  {\it General second order scalar-tensor theory, self tuning, and the Fab Four},
  Phys.\ Rev.\ Lett.\  {\bf 108}, 051101 (2012),
  [arXiv:1106.2000 [hep-th]].
  %%CITATION = doi:10.1103/PhysRevLett.108.051101;%%
  %294 citations counted in INSPIRE as of 16 Dec 2019
	
	%\cite{Charmousis:2011ea}
\bibitem{Charmousis:2011ea} 
  C.~Charmousis, E.~J.~Copeland, A.~Padilla and P.~M.~Saffin,
  {\it Self-tuning and the derivation of a class of scalar-tensor theories},
  Phys.\ Rev.\ D {\bf 85}, 104040 (2012),
  [arXiv:1112.4866 [hep-th]].
  %%CITATION = doi:10.1103/PhysRevD.85.104040;%%
  %109 citations counted in INSPIRE as of 16 Dec 2019

	%\cite{Starobinsky:2016kua}
\bibitem{Starobinsky:2016kua} 
  A.~A.~Starobinsky, S.~V.~Sushkov and M.~S.~Volkov,
  {\it The screening Horndeski cosmologies},
  JCAP {\bf 1606}, 007 (2016),
  [arXiv:1604.06085 [hep-th]].
  %%CITATION = doi:10.1088/1475-7516/2016/06/007;%%
  %30 citations counted in INSPIRE as of 16 Dec 2019
	
	%\cite{Bruneton:2012zk}
\bibitem{Bruneton:2012zk} 
  J.~P.~Bruneton, M.~Rinaldi, A.~Kanfon, A.~Hees, S.~Schlogel and A.~Fuzfa,
  {\it Fab Four: When John and George play gravitation and cosmology},
  Adv.\ Astron.\  {\bf 2012}, 430694 (2012),
  [arXiv:1203.4446 [gr-qc]].
  %%CITATION = doi:10.1155/2012/430694;%%
  %43 citations counted in INSPIRE as of 16 Dec 2019


 %\cite{Hui:2012qt}
  \bibitem{Hui:2012qt} 
  L.~Hui and A.~Nicolis,
  {\it No-Hair Theorem for the Galileon},
  Phys.\ Rev.\ Lett.\  {\bf 110}, 241104 (2013),
  [arXiv:1202.1296 [hep-th]].
  %%CITATION = doi:10.1103/PhysRevLett.110.241104;%%
  %109 citations counted in INSPIRE as of 28 Dec 2018

	\bibitem{Bravo-Gaete:2013dca} 
  M.~Bravo-Gaete and M.~Hassaine,
  {\it Lifshitz black holes with a time-dependent scalar field in a Horndeski theory},
  Phys.\ Rev.\ D {\bf 89}, 104028 (2014),
  [arXiv:1312.7736 [hep-th]].
  %%CITATION = doi:10.1103/PhysRevD.89.104028;%%
  %39 citations counted in INSPIRE as of 31 Dec 2018
  
  \bibitem{Babichev:2013cya} 
  E.~Babichev and C.~Charmousis,
  {\it Dressing a black hole with a time-dependent Galileon},
  JHEP {\bf 1408}, 106 (2014),
  [arXiv:1312.3204 [gr-qc]].
  %%CITATION = doi:10.1007/JHEP08(2014)106;%%
  %148 citations counted in INSPIRE as of 31 Dec 2018


%\cite{Fischler:2012ca}
\bibitem{Fischler:2012ca}
W.~Fischler and S.~Kundu,
{\it Strongly Coupled Gauge Theories: High and Low Temperature Behavior of Non-local Observables},
JHEP \textbf{05}, 098 (2013),
[arXiv:1212.2643 [hep-th]].
%90 citations counted in INSPIRE as of 04 Jan 2022


%\cite{Gioev:2006zz}
\bibitem{Gioev:2006zz}
D.~Gioev and I.~Klich,
{\it Entanglement Entropy of Fermions in Any Dimension and the Widom Conjecture},
Phys. Rev. Lett. \textbf{96}, 100503 (2006),
[arXiv:quant-ph/0504151 [quant-ph]].
%198 citations counted in INSPIRE as of 05 Jan 2022

%\cite{Wolf:2006zzb}
\bibitem{Wolf:2006zzb}
M.~M.~Wolf,
{\it Violation of the entropic area law for Fermions},
Phys. Rev. Lett. \textbf{96}, 010404 (2006),
[arXiv:quant-ph/0503219 [quant-ph]].
%208 citations counted in INSPIRE as of 05 Jan 2022

\end{thebibliography}
\end{document}